\documentclass[%
    superscriptaddress,
    preprint,
    preprintnumbers,
    amsmath,amssymb,
    aps,
]{revtex4-1}
\usepackage{graphicx}
\usepackage[osf]{mathpazo}
\usepackage{enumerate}
\usepackage[caption=false]{subfig}
\usepackage[monochrome]{xcolor}
\usepackage{color}
\usepackage{ulem}

\begin{document}
\preprint{KEK-TH-2291}
\preprint{J-PARC-TH-0237}
\title{Application of the Uniformized Mittag-Leffler Expansion to $\Lambda(1405)$}

\author{Wren A. Yamada}
\email{wren-phys@g.ecc.u-tokyo.ac.jp}
\affiliation{Department of Physics, Faculty of Science, University of Tokyo, 7-3-1 Hongo Bunkyo-ku Tokyo 113-0033, Japan}%
\affiliation{Theory Center, Institute of Particle and Nuclear Studies (IPNS), High Energy Accelerator Research Organization (KEK), 1-1 Oho, Tsukuba, Ibaraki, 205-0801, Japan}%

\author{Osamu Morimatsu}
\email{osamu.morimatsu@kek.jp}
\affiliation{Department of Physics, Faculty of Science, University of Tokyo, 7-3-1 Hongo Bunkyo-ku Tokyo 113-0033, Japan}%
\affiliation{Theory Center, Institute of Particle and Nuclear Studies (IPNS), High Energy Accelerator Research Organization (KEK), 1-1 Oho, Tsukuba, Ibaraki, 205-0801, Japan}%
\affiliation{Department of Particle and Nuclear Studies,
Graduate University for Advanced Studies (SOKENDAI),
1-1 Oho, Tsukuba, Ibaraki 305-0801, Japan}%
\date{\today}
\begin{abstract}
We study the pole properties of $\Lambda(1405)$ in a model-independent manner by applying the Uniformized Mittag-Leffler expansion proposed in our previous paper.
The resonant energy, width and residues {\color{blue} are determined} by expanding the observable as a sum of resonant-pole pairs {\color{blue} under an appropriate parameterization} which expresses the observable {\color{blue} to be single-valued}, and fitting it to experimental data{\color{blue} of} the invariant-mass distribution of $\pi^+\Sigma^-$, $\pi^-\Sigma^+$, $\pi^0\Sigma^0$ final states in the reaction, $\gamma p \rightarrow K^+ \pi \Sigma$, and the elastic and inelastic cross section, $K^-p\to K^-p$, $\bar{K}^0n$, $\pi^+\Sigma^-$, $\pi^-\Sigma^+$.
As we {\color{blue} gradually} increase the number of pairs from one to three, the first pair converges while the second and third pairs emerge {\color{blue} further and further} away from the first pair, 
{\color{blue}implying} that the Uniformized Mittag-Leffler expansion with three pairs is almost convergent in the vicinity of the $\Lambda(1405)$.
The broad peak structure between the $\pi\Sigma$ and $\bar{K} N$ thresholds regarded to be $\Lambda(1405)$ is explained by {\color{blue} a single} pair with {\color{blue}a} resonant energy {\color{blue}of} 1420 $\pm$ 1 MeV, and {\color{red} a half} width of 48 $\pm$ 2 MeV{\color{red},
which is consistent with the single-pole picture of $\Lambda(1405)$}.
{\color{blue} We conclude that} the Uniformized Mittag-Leffler expansion turns out to be {\color{blue} a} very powerful {\color{blue} method to obtain resonance energy, width and residues from the near-threshold spectrum}.
\par
\end{abstract}
\maketitle
\section{Introduction}\label{sec:intro}
In recent years many hadron resonances, in particular, candidates of exotic hadrons have been found near the thresholds of hadronic channels {\color{red} \cite{Guo:2017jvc,karliner}}.
Due to the threshold effects, their spectra are significantly distorted from the Breit-Wigner form {\color{red}\cite{Breit:1936zzb}},
{\color{blue}
\begin{equation}\label{eq:Breit-Wigner}
  \mathcal{A}(\sqrt{s}) \propto \frac{\Gamma_R}{\sqrt{s}-M_R-i\Gamma_R/2},
\end{equation}}
{\color{blue}making} it challenging to extract information of resonances such as the resonance {\color{blue}energy} and width from the observed spectra {\color{blue} in a model-independent manner}. 
\par
In our previous paper \cite{PhysRevC.102.055201}, we proposed the Uniformized Mittag-Leffler expansion approach, a model-independent approach that incorporates the resonant and threshold behaviors appropriately.
We showed that when choosing an appropriate parameterization {\color{red}\cite{Newton:book,KATO1965130}}, the $S$-matrix is a meromorphic function and can be expressed by the Mittag-Leffler expansion {\color{red}\cite{HUMBLET1961529,ROMO197861,BANG1978381,BERGGREN1982261}}. It is explicitly written by the positions and residues of the bound and resonant poles.
The symmetry condition of the pole properties of the $S$-matrix forces the series to obey the proper threshold behaviors.
Following our proposition, we demonstrated the method by using data of a double-channel model calculation, with {\color{blue} isospin} $I=0$, $\overline{K}N$ and $\pi\Sigma$ channels.
\par
The next step would naturally be the demonstration of the method to actual hadronic spectra. 
In the present paper, we apply the Uniformized Mittag-Leffler expansion to experimental data of the spectrum around $\Lambda(1405)$; a resonance situated between the $\pi\Sigma$ and $\bar{K}N$ threshold {\color{red}\cite{Dalitz:1959dn,DALITZ1959100}}, which has been a topic of interest in studies involving baryons with strangeness {\color{red}\cite{mai2018status,KAMIYA201641,CIEPLY201617}}.  
It has been naturally described as a hadronic molecular state generated from hadronic degrees of freedom {\color{red}\cite{Guo:2017jvc}}.
while hardly interpreted as an excitation in the standard three-quark description. 
Moreover, calculations in the chiral-unitary model, such as {\color{red}\cite{JIDO2003181,Oller:2000fj,PhysRevC.77.035204,Hyodo:2011ur,IKEDA201163,IKEDA201298}}, display a double-pole structure in the region of $\Lambda(1405)$,
{\color{blue}contrary to} phenomenological local potential models, such as {\color{red}\cite{Akaishi:2010wt,Myint:2018ypc,Revai:2017isg,Revai:2019ipq}}, {\color{blue}which} predict a single-pole structure.
{\color{blue}In order to settle the debate between the single-pole or double-pole structure of $\Lambda(1405)$, a model-independent analysis is strongly in need.}
Owing to these circumstances, the $\Lambda(1405)$ resonance serves as an ideal target for the application of our method.
\par
We apply the Uniformized Mittag-Leffler expansion approach to the scattering processes of $K^-p$ elastic and inelastic cross sections \cite{Abrams:1965,Bangerter:1980px,Ciborowski:1982et,Csejthey-Barth:1965izu,Humphrey:1962,Mast:1975pv,Sakitt:1965}, and the invariant-mass distributions of $\pi\Sigma$ final states in the reaction, $\gamma p \rightarrow K^+ \pi \Sigma$, measured with CLAS at Jefferson Lab \cite{Moriya:2013eb}.
Under the assumption that the spectrum is dominantly generated from the coupled-channel dynamics of the $\pi\Sigma$-$\bar{K}N$ system, we determine the resonance energy, width, and residues of $\Lambda(1405)$ in a model-independent manner.
\section{Uniformized Mittag-Leffler expansion approach}\label{sec:uni_ml_approach}
Here we will briefly review our new approach proposed in Ref.\,\cite{PhysRevC.102.055201} for a better understanding of its application to actual experimental data in the following section, and to clarify our conventions.
\par
From the Cauchy integration principle, a meromorphic function $f(z)$ can be written as, 
\begin{equation}\label{eq:contour}
  f(z)=\frac{1}{2\pi i}\oint_{\gamma} dw \frac{f(w)}{w-z}+\sum_{n}\frac{c_{n}}{z-z_{n}},
\end{equation}
where $z_n$, $c_n$ are the poles and residues of $f$, and $\gamma$ is a circular contour around the origin with a radius taken to infinity.
If the integral term in Eq.\,(\ref{eq:contour}) vanishes as we take the radius of $\gamma$ to infinity, the meromorphic function $f$ can be written as,
\begin{equation}\label{eq:mt}
  f(z)=\sum_n\frac{c_n}{z-z_n},
\end{equation}
which is a Mittag-Leffler expansion of $f$.
Note that this form is explicitly written by a simple series of the pole position and the residue of $f$.
\par
When the integral term in Eq.\,(\ref{eq:contour}) does not vanish, or diverges, we can always consider a subtraction, so that the integral takes a form with better convergence to zero.
For example, let us consider $g(z)=(f(z)-f(0))/z$ instead of $f(z)$ in Eq.\,(\ref{eq:contour}).
If the integral term in Eq.\,(\ref{eq:contour}) for $g(z)$ vanishes, the once-subtracted form of Eq.\,(\ref{eq:mt}) can be written as,
\begin{equation}\label{eq:mt_sub}
  f(z)=f(0)+\sum_n\biggl[\frac{c_n}{z-z_n}+\frac{c_n}{z_n}\biggr],
\end{equation}
which differs from Eq.\,(\ref{eq:mt}) by a constant that corresponds to the subtraction.
\par
Now, let us consider a two-body system. Observables, such as two-body cross sections, $\sigma$, or the distributions of two-body final states with invariant mass, $M$, in some reaction (e.g.\,$\pi\Sigma$ final states in the reaction, $\gamma p \rightarrow K^+ \pi \Sigma$), $d\sigma/dM$, are related to the $T$-matrix, $\mathcal{T}$, and Green's function, $\mathcal{G}$ as {\color{red}\cite{FetterWalecka,BERTSCH1975125,MORIMATSU1994679}},
\begin{equation}
  \sigma=\frac{1}{16\pi^2s}\frac{k_f}{k_i}\text{Im }\mathcal{T}, \\
\end{equation}
or
\begin{equation}
  \frac{d\sigma}{dM}=\text{Im }\mathcal{F^\dagger G F} \label{eq:crs_analit},
\end{equation}
where $s$ is the {\color{red} center-of-mass} energy squared, $k_f$, $k_i$ are the final and initial momenta in the {\color{red} center-of-mass} frame, repectively, and $\mathcal{F}$ represents the vertex producing two-body final states.
For our convenience let $\mathcal{A}$ represent either $\mathcal{T}$ or $\mathcal{F^\dagger G F}$. $\mathcal{A}$ has the same analytic structure as the $S$-matrix.
\par
From the unitary condition, the $S$-matrix has a branch cut running from each threshold along the positive real axis in the $\sqrt{s}$-plane to infinity, known as \textit{unitary cuts}.
Thus, $\mathcal{A}$ is not meromorphic, and Eq.\,(\ref{eq:mt}) or Eq.\,(\ref{eq:mt_sub}) cannot be applied directly. 
To explicitly write $\mathcal{A}$ in the form of a Mittag-Leffler expansion, one must choose an appropriate parameterization so that $\mathcal{A}$ becomes meromorphic. This process is called \textit{uniformization}. Once uniformization is performed, $\mathcal{A}$ can be decomposed into a series of the form of Eq.\,(\ref{eq:mt}) or Eq.\,(\ref{eq:mt_sub}).
In addition, the unitarity of the $S$-matrix also imposes a symmetry condition 
on the position of the pole positions and the residues of $\mathcal{A}$.
The poles are positioned symmetric about the imaginary axis in the uniformized $z$-plane, and the residues, $c_n(z_n)$, satisfy the following relationship,
\begin{equation}
    c_n(z_n)=-c_n^\ast(-z_n^\ast).
\end{equation}
Note that when considering the pole symmetry condition, the subtraction constant in Eq.\,(\ref{eq:mt_sub}) is real, and thus the imaginary part of Eq.\,(\ref{eq:mt}) and Eq.\,(\ref{eq:mt_sub}) take the same form.
\par
To summarize, by an appropriate choice of variable, $z$, the imaginary part of $\mathcal{A}$ can be written as,
\begin{equation}
  \text{Im~}\mathcal{A}(z)=\text{Im~}\sum_{n}\biggl(\frac{c_n}{z-z_n}-\frac{c_n^\ast}{z+z_n^\ast}\biggr),
\end{equation}
which we will call the Uniformized Mittag-Leffler expansion.
Expressing observables in the form of the Uniformized Mittag-Leffler expansion and comparing them with the actual experimental data, we can obtain the pole positions and residues of the observables from experimental data in a model-independent manner.
\textcolor{red}{Let us call this the Uniformized Mittag-Leffler expansion approach.}
\par
Explicit procedures are as follows:
\begin{enumerate}[i]
  \item Find an appropriate kinetic variable, $z$, which uniformizes the $\mathcal{S}$-matrix.
  \item Truncate the Uniformized Mittag-Leffler expansion, and approximate $\mathcal{A}(z)$ by a few ($m$) pairs of the pole terms as,
  {\color{blue}
  \begin{equation}\label{eq:mittag_leffler_approach}
    \text{Im }\mathcal{A}(z)=\text{Im}\sum_{n=1}^m \left(\frac{c_n^{(m)}}{z-z_n^{(m)}} - \frac{c^{(m)\ast}_n}{z+z^{(m)\ast}_n}\right)
  \end{equation}}
  \item Determine the complex pole positions, {\color{blue}$z_n^{(m)}$}, and residues, {\color{blue}$c_n^{(m)}$}, ($n=1,\cdots, m$), by fitting $\mathcal{A}$ to the experimental data.
  \item {\color{blue}Regard converged $z_n^{(m)}$, $c_n^{(m)}$ as the actual pole positions and residues.}
\end{enumerate}
\par
The two-body double-channel $\mathcal{S}$-matrix can be expressed as a four-sheeted Riemann surface with unitary cuts running from each threshold $\epsilon_1$, $\epsilon_2$ to $\infty$ along the real axis, in the parameterization of {\color{red} center-of-mass} energy, $\sqrt{s}$.
The threshold energy $\epsilon_i$ is,
\begin{equation}
  \epsilon_i=M_i+m_i,
\end{equation}
where $M_i$, and $m_i$ are the masses of the two particles in channel $i=1$, $2$.
The four sheets in the $\sqrt{s}$-plane can be uniquely labeled by the sign of the imaginary part of $q_1$ and $q_2$, given by,
\begin{equation}
  q_i=\sqrt{s-\epsilon_i^2},
\end{equation}
which has a one-to-one correspondance with channel momenta. In this paper we label the four sheets by,
$(\sigma(q_1)~\sigma(q_2))$ where, 
\begin{equation}
  \sigma(q_i)=\text{sgn}(\text{Im }q_i).
\end{equation}
The physical sheet corresponds to sheet $(++)$.
\par
\begin{figure}[!htb]
  \centering
  \subfloat[]{\includegraphics[width=0.5\linewidth]{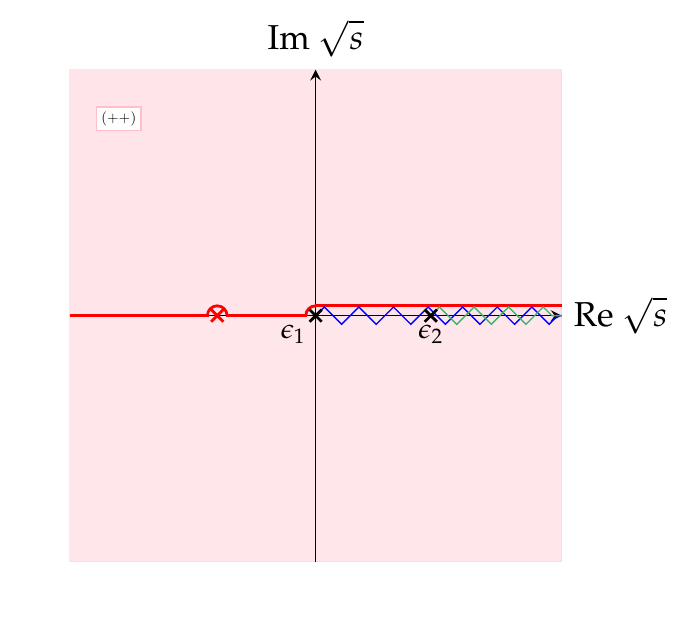}\label{fig:analytic_s_plane}}
  \subfloat[]{\includegraphics[width=0.5\linewidth]{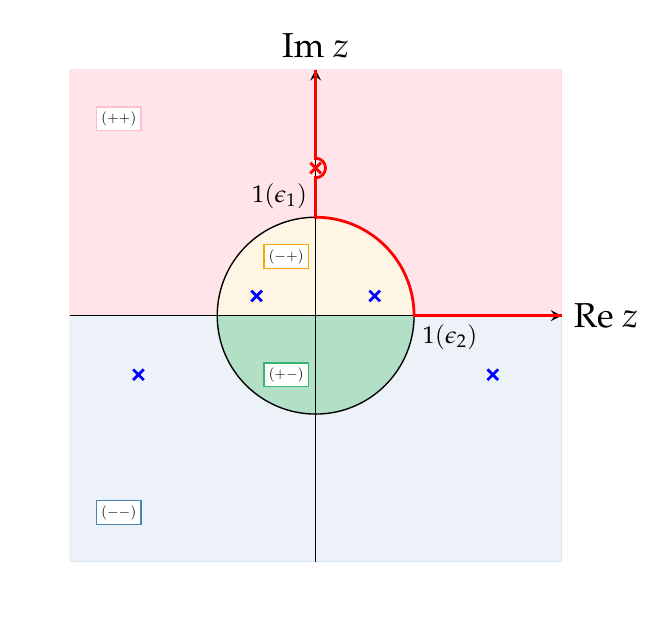}\label{fig:double_z}}
  \caption{Analytic structure of the double-channel $\mathcal{S}$-matrix in the $\sqrt{s}$-plane, Fig.\,(a), and the uniformized $z$-plane, Fig.\,(b). In the $\sqrt{s}$-plane, the unitary cuts run along the real axis from $\epsilon_1$ to $\infty$ (blue) and from $\epsilon_2$ to $\infty$ (green), and the four Riemann sheets of the $\sqrt{s}$-plane correspond to each region in $z$ labeled as $(\pm\pm)$. The red line shows the physical region accessible by experiment.}
\end{figure}

By the parameterization $z$ \cite{KATO1965130},
\begin{equation*}
  z=\frac{1+\sqrt{u}}{1-\sqrt{u}},\label{eq:z}
\end{equation*}
{\color{red}
where
\begin{equation}
  u=\frac{q_1-\Delta}{q_1+\Delta} \quad \text{and} \quad \Delta=\sqrt{\epsilon_2^2-\epsilon_1^2},
\end{equation}
}
the four-sheeted Riemann surface can be uniformized into a single complex plane so that $\mathcal{S}(z)$ is meromorphic.
The correspondance between the $\sqrt{s}$-plane and $z$-plane are shown in Fig.\,\ref{fig:double_z}.
The two thresholds, $\sqrt{s}=\epsilon_1$ and $\sqrt{s}=\epsilon_2$ are transformed to points on the unit circle $z=i$, and $z=1$, respectively. The imaginary axis above $i$, the unit circle between $i$ and $1$, and the real axis above $1$ correspond to
the physically accessable region of $\sqrt{s}<\epsilon_1$, $\epsilon_1<\sqrt{s}<\epsilon_2$, and $\sqrt{s}<\epsilon_2$, respectively.
\begin{figure}[htpb] 
  \centering
  \begin{tabular}{cc}
      \begin{minipage}{0.50\hsize}
          \centering
          \includegraphics[width=\linewidth]{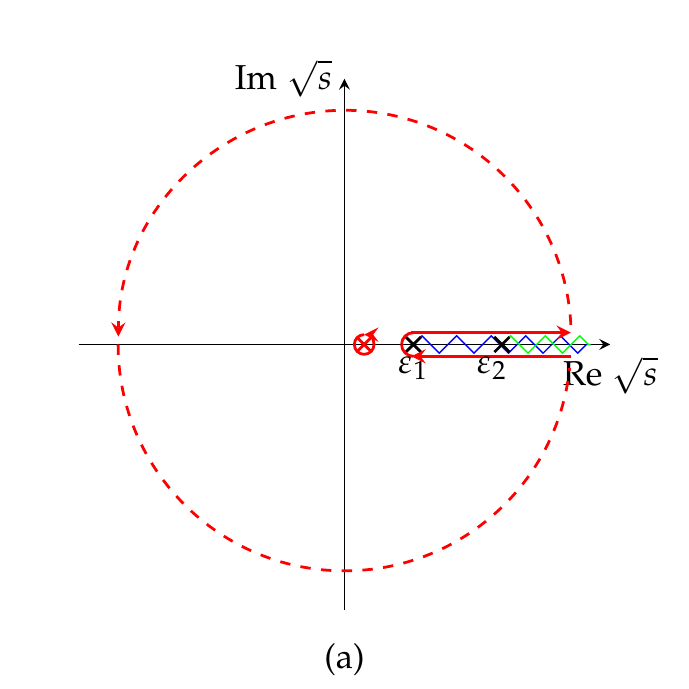}
      \end{minipage}
      \begin{minipage}{0.50\hsize}
          \centering
          \includegraphics[width=\linewidth]{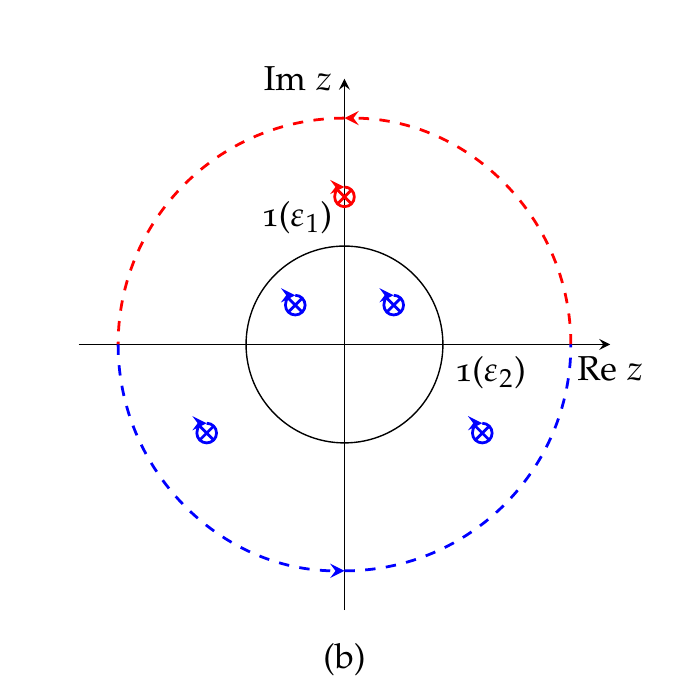}
      \end{minipage}
  \end{tabular}
  \caption{Schematic figure of the spectral representation, Fig.\,(a), and the Uniformized Mittag-Leffler expansion, Fig.\,(b), of the $\mathcal{S}$-matrix in the case of a double-channel two-body system.
  $\epsilon_1$, $\epsilon_2$ are the threshold energies of the two channels. The continuum contribution along the unitary cuts in Fig.\,(a), can be decomposed into resonant contributions from poles in the unphysical domain (blue) in Fig.\,(b).}
\end{figure}
\par
The contribution of a single resonant-pole pair, $\mathcal{A}_n$, is given in the vicinity of $\sqrt{s}=\epsilon_1$ as,
    \begin{align}
      \textrm{Im } \mathcal{A}_n(z) =
      \begin{cases}
          0, & (\sqrt{s} < \epsilon_1)\\
          \displaystyle{-\textrm{Im}\frac{2c_n}{(z_n-i)^2}\frac{k_1}{\Delta}+\mathcal{O}(k_1^2)}, & (\sqrt{s} > \epsilon_1)
      \end{cases}
      \label{eq:thr_10}
  \end{align}
   and in the vicinity of $\sqrt{s}=\epsilon_2$ as,
     \begin{align}
      \textrm{Im } \mathcal{A}_n(z) = 
      \begin{cases}
          \displaystyle{\textrm{Im}\frac{2c_n}{1-z_n^2}-\textrm{Re}\frac{4c_nz_n}{(1-z_n^2)^2}\frac{\tilde{k}_2}{\Delta}+\mathcal{O}(\tilde{k}_2^2)}, &  (\sqrt{s} < \epsilon_2) \\
           \displaystyle{\textrm{Im}\frac{2c_n}{1-z_n^2}-\textrm{Im}\frac{2c_n(1+z_n^2)}{(1-z_n^2)^2}\frac{k_2}{\Delta}+\mathcal{O}(k_2^2)}, &  (\sqrt{s} > \epsilon_2)
      \end{cases}
    \label{eq:thr_20}
    \end{align}
where $k_1$ and $k_2$ are the momenta in channel 1 and 2, respectively, and $\tilde{k}_2$ is defined by $k_2=i\tilde{k}_2$. 
Eqs.\,(\ref{eq:thr_10}) and (\ref{eq:thr_20}) describe the proper threshold behaviors.
{\color{red} Therefore, we will always take into account pairs of poles together in the Uniformized Mittag-Leffler expansion.
It should be noted, however, that the conjugate poles do not affect the the structure of resonances well above the lowest threshold, because they are more distant as the energy becomes higher above the lowest threshold.}
\par
If a pole is located close to the physical region and sufficiently away from the threshold, its contribution is approximately given by Eq.\,(\ref{eq:Breit-Wigner}) with a complex residue as,
\begin{align}
  {\rm Im} \frac{c_n}{z-z_n} &\approx  {\rm Im} \frac{\tilde{c}_n}{\sqrt{s}-{\sqrt{s}_n}}  \nonumber\\
  &= A \cos \theta ~ \frac{\Gamma_n/2}{\left(\sqrt{s}-{\epsilon_n}\right)^2+\Gamma_n^2/4 } + A\sin \theta ~\frac{\sqrt{s}-{\epsilon_n}}{\left(\sqrt{s}-{\epsilon_n}\right)^2+\Gamma_n^2/4 }\label{eq:breit_wigner_w_phase},
\end{align}
where {\color{red} $\sqrt{s}_n=\epsilon_n-i\Gamma_n/2$ and $\tilde{c}_n=c_n\left[dz/d\sqrt{s}\right]^{-1}_{\sqrt{s}=\sqrt{s}_n}=Ae^{i\theta}$ are, respectively, the position and residue of the pole in the parameterization, $\sqrt{s}$, corresponding to $z_n$}.
The standard Breit-Wigner form corresponds to the particular case of $\theta=0$.
Note that the approximation in the first line of Eq.\,(\ref{eq:breit_wigner_w_phase}) only holds for narrow resonances distant from the threshold. 
On some local coordinate system, the mapping between $\sqrt{s}$ and $z$ is a conformal map, thus preserving the local geometric structure, meaning when $|z-z_n|$ is small and away from critical points, $z=i,1$, $1/(z-z_n)\approx\alpha/(\sqrt{s}-\sqrt{s}_n)$.
{\color{red} In the neighbourhood of the thresholds, or in the region of large $\Gamma$,} the mapping between $\sqrt{s}$ and $z$ is warped significantly, so that the approximation breaks down.
\clearpage
{\color{blue}\section{Application to the experimental spectrum of $\Lambda(1405)$}\label{sec:lamda1405}}
We now apply our method to the {\color{blue} experimental spectrum of} $\Lambda(1405)$, regarding $\Lambda(1405)$ as a resonance in the coupled $I=0$ two channels, $\pi\Sigma$ and $\bar{K}N$.
\subsection{Fitting procedure}
We fit the Uniformized Mittag-Leffler expansion with $m$ resonant-pole pairs to the invariant-mass distributions of $\pi^+\Sigma^-$, $\pi^-\Sigma^+$ and $\pi^0\Sigma^0$ final states in the reaction, $\gamma p \rightarrow K^+ \pi \Sigma$, measured with CLAS at Jefferson Lab for center-of-mass energies $1.95 < W < 2.85$ GeV \cite{Moriya:2013eb} as
{\color{blue}
\begin{align}\label{eq:fit_pole3}
    \frac{d\sigma^{W}}{dM}&=\text{Im}\sum_{n=1}^m \left(\frac{c_n^{W(m)}}{z-z_n^{(m)}} - \frac{c^{W(m)\ast}_n}{z+z^{(m)\ast}_n}\right),
\end{align}}
and the $K^-p$ elastic and inelastic cross sections, $K^-p\to K^-p$, $\bar{K}^0n$, $\pi^+\Sigma^-$, $\pi^-\Sigma^+$, \cite{Abrams:1965,Bangerter:1980px,Ciborowski:1982et,Csejthey-Barth:1965izu,Humphrey:1962,Mast:1975pv,Sakitt:1965} as
{\color{blue}
\begin{align}\label{eq:fit_pole3}
  \sigma^{if}&=\frac{1}{16\pi^2s}\frac{k_f}{k_i}\text{Im}\sum_{n=1}^m \left(\frac{c_n^{if(m)}}{z-z_n^{(m)}} - \frac{c^{if(m)\ast}_n}{z+z^{(m)\ast}_n}\right).
\end{align}}
In Eq.\,(17), $d\sigma^{W}/dM$ is the distribution of the $\pi\Sigma$ invariant-mass, $M$, with the center-of-mass energy, $W$, of the reaction, $\gamma p \rightarrow K^+ \pi \Sigma$.
In Eq.\,(18), $\sigma^{if}$ is the cross section of the scattering process, $i\to f$, $s$ is the center-of-mass energy squared and $k_i$ ($k_f$) is the momentum of the initial (final) state in the center-of-mass frame.
{\color{red} The invariant mass distribution was measured in 9 different {\color{red} center-of-mass} energies, $W$, in the range of 1.95-2.85 GeV for each channel, $\pi^+\Sigma^-$, $\pi^-\Sigma^+$, and $\pi^0\Sigma^0$.
Each dataset of $d\sigma^{W}/dM$ and $\sigma^{if}$ is fitted with different residues but common pole positions.
Therefore, in the case of $m$ resonant-pole pairs and $N$ data sets we have $m$ and $mN$ complex parameters for the pole positions and residues, respectively.}  
The behavior of the $\pi\Sigma$ invariant-mass distributions in the reaction, $\gamma p \rightarrow K^+ \pi \Sigma$, is sensitive to the $\pi\Sigma$ threshold energies, which are slightly different for the $\pi^+\Sigma^-$, $\pi^-\Sigma^+$ and $\pi^0\Sigma^0$ channels.
Therefore, we take into account the difference of the threshold energies with minimum modifications, though we basically regard $\Lambda(1405)$ as a resonance in the coupled two channels, $\pi\Sigma$ and $\bar{K}N$ with isospin as an approximately good quantum number.
Namely, we define $z$ differently for each of the $\pi^+\Sigma^-$, $\pi^-\Sigma^+$ and $\pi^0\Sigma^0$ channels with slightly different $\pi\Sigma$ threshold energies in the fit of the $\pi\Sigma$ invariant-mass distributions.
We neither take into account the difference of $\bar K N$ threshold energies in the $\pi\Sigma$ invariant-mass distributions nor the difference of $\pi\Sigma$ and $\bar K N$ threshold energies in the $K^-p$ elastic and inelastic cross sections because it is simply unnecessary.
{\color{red} As explained above, each dataset of $d\sigma^{W}/dM$ and $\sigma^{if}$} is fitted with different residues but common pole positions such that all pole positions are common on the $\sqrt{s}$-plane.
This means that the pole positions on the $z$ plane are slightly different for the $\pi^+\Sigma^-$, $\pi^-\Sigma^+$ and $\pi^0\Sigma^0$ invariant-mass distributions.
When we show the pole positions in the $z$-plane, $z$ is defined for the $\pi^+\Sigma^-$ channel.
The differences, however, are small and will be ignored in the following discussions.
\par
\begin{figure}[htpb]
  \centering
  \includegraphics[width=0.4\linewidth]{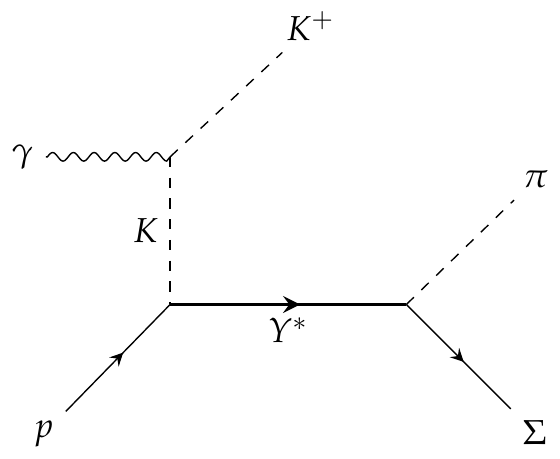}
  \caption{A schematic diagram of the process${\color{red}, \lq}\gamma p \to K^+Y^\ast \to K^+\pi\Sigma{\color{red}\rq,}$ measured in the CLAS experiment \cite{Moriya:2013eb}.}
\end{figure}
\par
We start from one resonant-pole pair, $m=1$, and {\color{blue} gradually} increase the number of pairs up to $m=3$.
The reduced chi square values are 5.74, 2.65, and 1.18 for cases, $m=1$, $2$ and $3$, respectively, and the case, $m=3$, best fitted the spectrum among them. 
We will present the results of the case, $m=3$, in detail in subsection {\color{blue}B.} and discuss the convergence of the Uniformized Mittag-Leffler expansion from $m=1$ to $m=3$ in subsection {\color{blue}C.}
\subsection{Results of Uniformized Mittag-Leffler expansion with $m=3$}\label{sec:result}
\begin{figure}[htpb] 
  \includegraphics[width=0.7\linewidth]{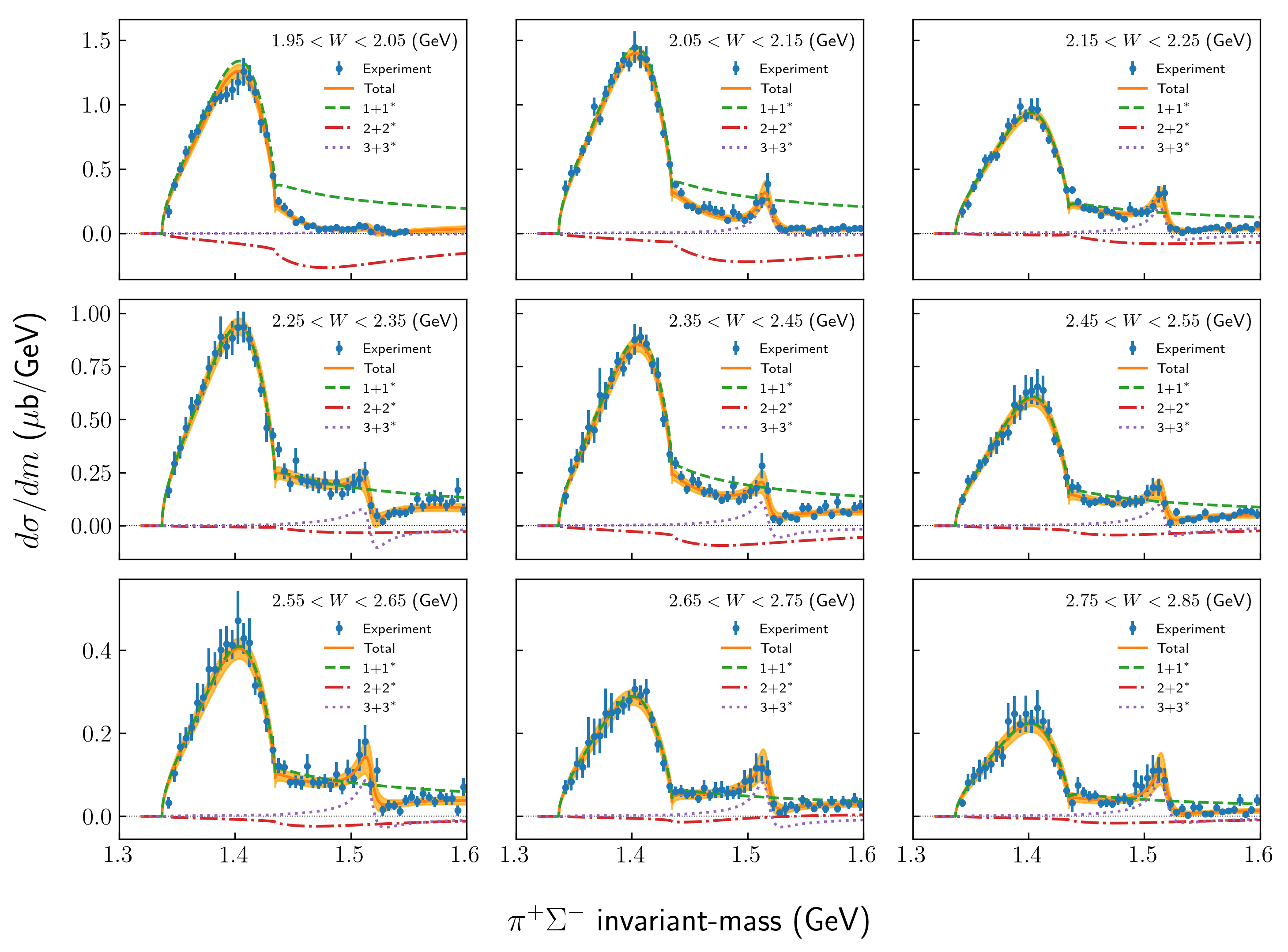}
  \caption{Results for the invariant-mass distributions of $\pi^+\Sigma^-$ in the reaction, $\gamma p \rightarrow K^+ \pi \Sigma$, in nine bins of the center-of-mass energy, $W$.
  The (blue) dotts with bars are the experimental data.
  The (orange) bands represent the $2\sigma$-confidence interval of the fit by the Uniformized Mittag-Leffler expansion with $m=3$.
  The (green) dashed, (red) dot-dashed and (purple) dotted lines represent the contributions from indivisual resonant-pole pairs of $1+1^*$, $2+2^*$ and $3+3^*$, respectively.}
  \label{fitting_curve_psm}
\end{figure}

\begin{figure}[htpb] 
  \centering
  \includegraphics[width=0.7\linewidth]{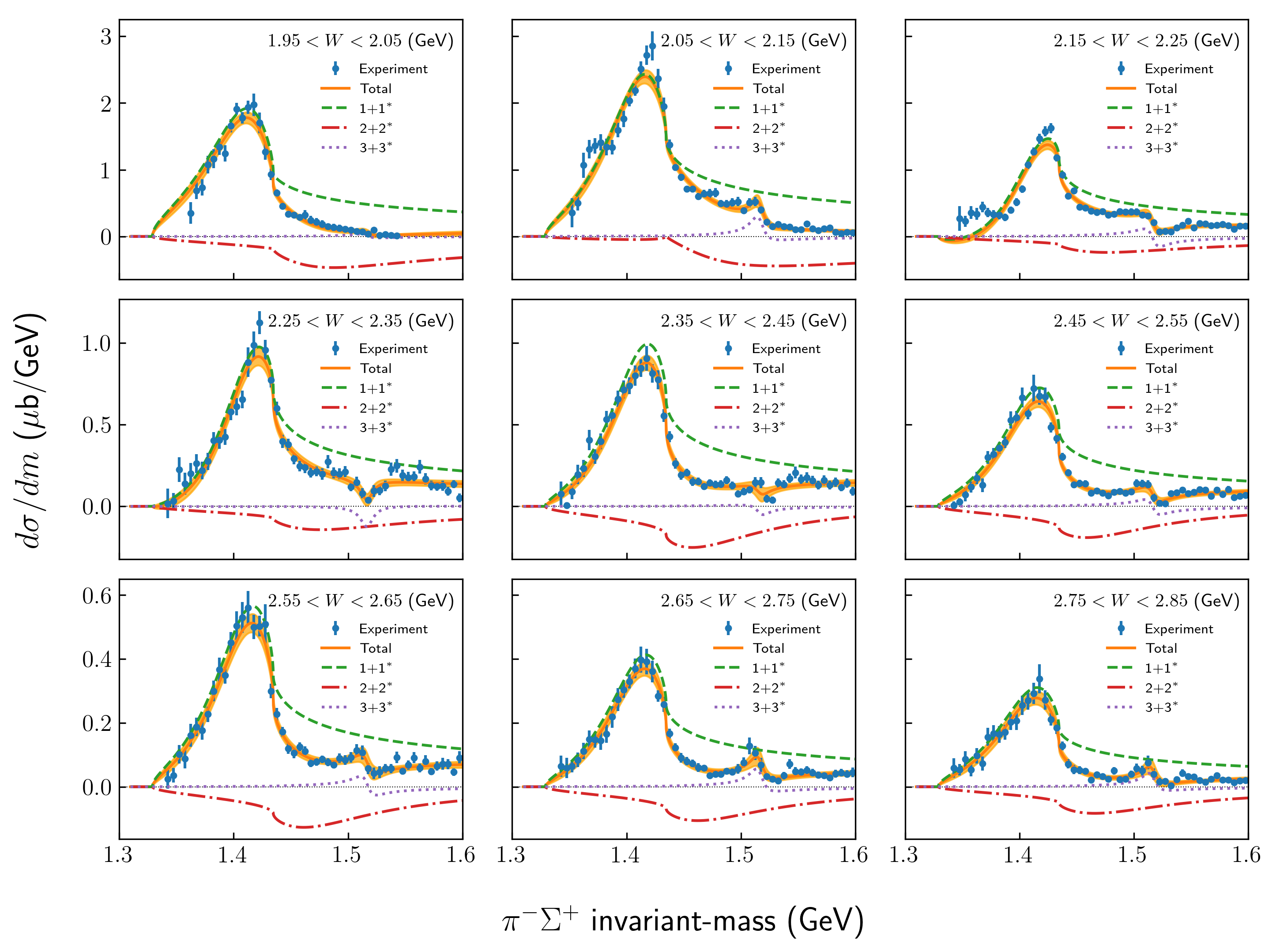}
  \caption{Results for the invariant-mass distributions of $\pi^-\Sigma^+$. Details are the same as in Fig. \ref{fitting_curve_psm}.}
  \label{fitting_curve_psp}
\end{figure}

\begin{figure}[htpb] 
  \centering
  \includegraphics[width=0.7\linewidth]{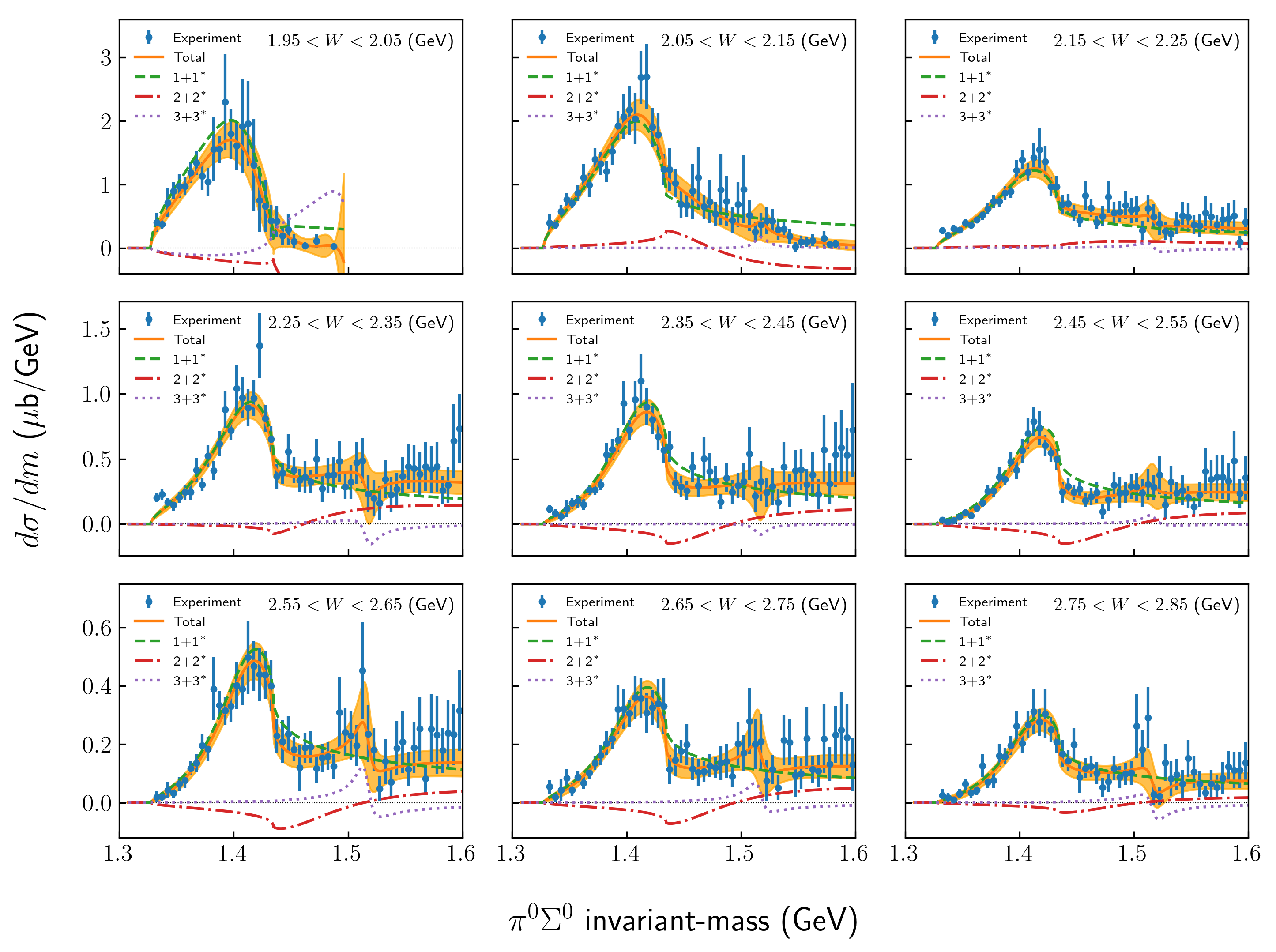}
  \caption{Results for the invariant-mass distributions of $\pi^0\Sigma^0$. Details are the same as in Fig.\,\ref{fitting_curve_psm}.
   {\color{red} In the $\pi^0\Sigma^0$ channel the $2\sigma$-confidence interval is wider than that in the $\pi^+\Sigma^-$ and $\pi^-\Sigma^+$ channels,
   due to large experimental errors.}}
  \label{fitting_curve_pso}
\end{figure}

\begin{figure}[htpb]
  \centering
  \includegraphics[width=0.7\linewidth]{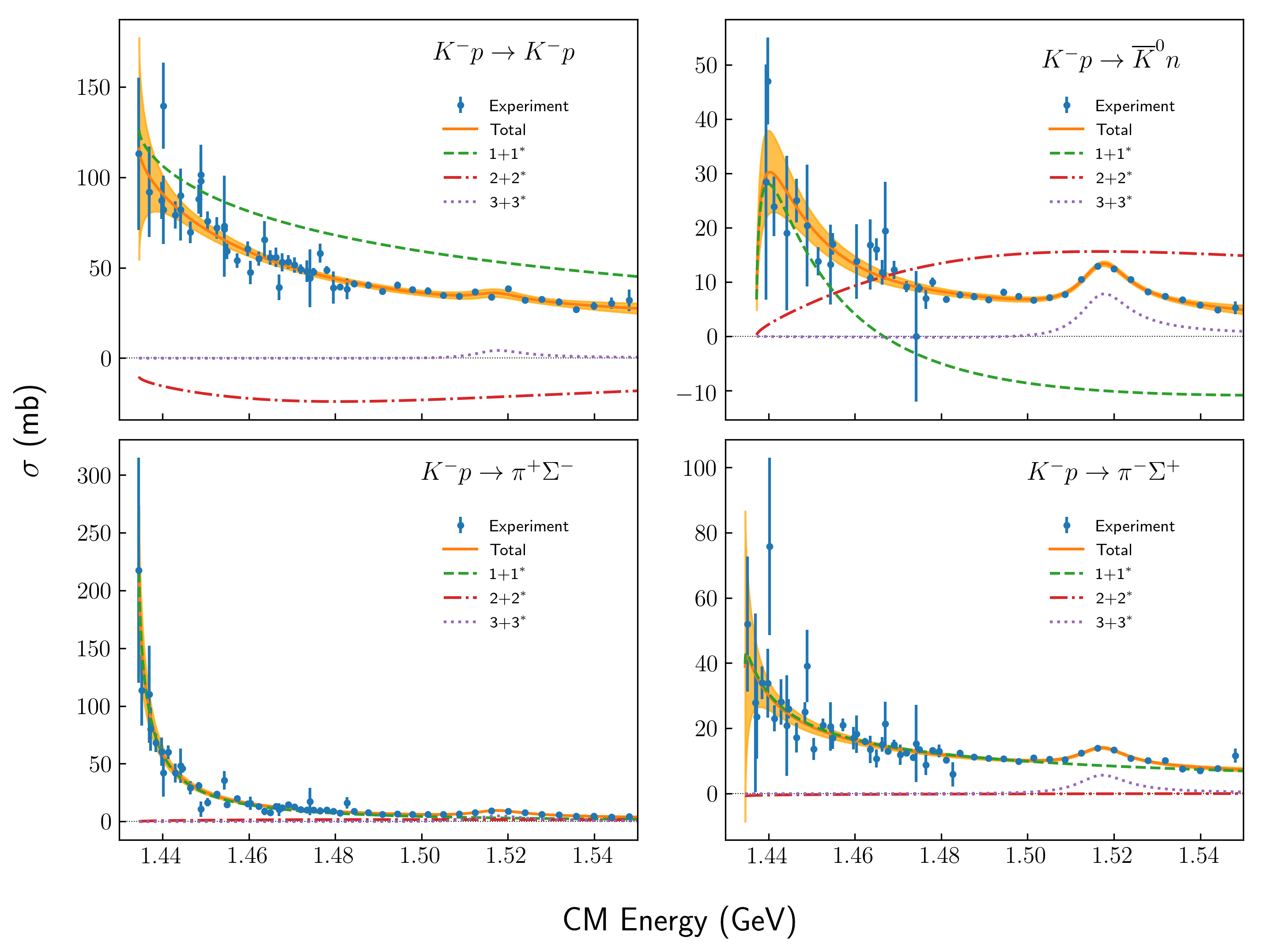}
  \caption{Results for the cross sections, $K^-p\to K^-p$, $\bar{K}^0n$, $\pi^+\Sigma^-$, $\pi^-\Sigma^+$. Details are the same as in Fig.\,\ref{fitting_curve_psm}.}
  \label{fitting_curve_scat}
\end{figure}

Figs.\,\ref{fitting_curve_psm}-\ref{fitting_curve_pso} show the fitted invariant-mass distributions of $\pi^+\Sigma^-$, $\pi^-\Sigma^+$, $\pi^0\Sigma^0$ in the reaction, $\gamma p \rightarrow K^+ \pi \Sigma$, and Fig.\,\ref{fitting_curve_scat} shows the elastic and inelastic cross sections, $K^-p\to K^-p$, $\bar{K}^0n$, $\pi^+\Sigma^-$, $\pi^-\Sigma^+$.
{\color{red} Since $m=3$ and $N=31$, we have $3$ and $93$ complex parameters for the pole positions and residues, respectively.} 
The Uniformized Mittag-Leffler expansion with $m=3$ fits experimental data very well,
which is confirmed also by the reduced chi-squared value, $1.18$.
The spectrum between the $\pi\Sigma$ and $\bar{K}N$ thresholds is mostly given by the resonant-pole pair $1+1^\ast$, while the spectrum above the $\bar{K}N$ threshold is given by the sum of $1+1^\ast$ and $2+2^\ast$ except for the narrow structure around 1520 MeV, which is explained by the contribution of $3+3^\ast$. The contribution of $3+3^\ast$ can be considered as the remnant of $\Lambda(1520)$, which was not exactly subtracted from the bare experimental data {\color{red}\cite{Moriya:2013eb}}.
Also, an extra structure is observed in the invariant-mass distributions of $\pi^-\Sigma^+$ around 1350 MeV. 
To explain such a structure, it may be necessary to consider contributions from additional resonant-pole pairs in the isospin $I=1$ sector.
\par
\begin{figure}[htpb]
  \centering
  \includegraphics[width=0.6\linewidth]{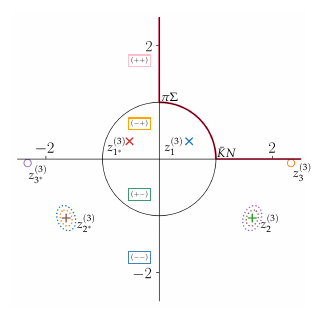}
  \caption{Results for the pole positions on the $z$-plane by the uniformized Mittag-Leffler expansion with $m=3$.
  Let us denote $z_{n^\ast}^{(m)}=-z_n^{(m)\ast}$.
  Two dotted lines around the poles represent the  70\% and 95\% confidence intervals of the position of the poles, respectively, from inside to outside.
  The (red) thick line represents the physical region accessible in the experiment
  and labels, $\pi\Sigma$ and $\bar{K}N$, represent the corresponding thresholds.
}  \label{pole_pos_z}
\end{figure}
\begin{figure}[htpb]
  \centering
  \includegraphics[width=0.8\linewidth]{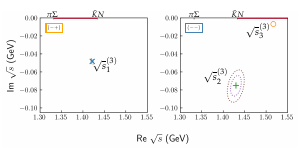}
  \caption{Results for the pole positions on the $\sqrt{s}$-plane. Details are the same as in Fig. \ref{pole_pos_z}}.
  \label{pole_pos_s}
\end{figure}

The positions of poles are tabulated in Tab.\,\ref{tab:pole} and are shown on the $z$-plane and $\sqrt{s}$-plane in Fig.\,\ref{pole_pos_z} and  Fig.\,\ref{pole_pos_s}, respectively.
In Fig.\,\ref{pole_pos_s} $(\text{sgn(Im}\,q_1)\,\text{sgn(Im}\,q_2))$ labels the four sheets of the $\sqrt{s}$-plane, where $q_1$ ($q_2$) corresponds to the relative momentum in the $\pi\Sigma$ ($\bar{K}N$) channel.
The sheet ($-\,+$) is located adjacent to the physical energy between the $\pi\Sigma$ and $\bar{K}N$ thresholds while the sheet ($-\,-)$  above the $\bar{K}N$ threshold.
Pole $1$ is positioned on the sheet ($-\,+$) right below the $\bar{K}N$ threshold at the complex energy of 1420-47i MeV.
Poles $2$ and $3$ are positioned on the sheet ($-\,-$), at the complex energies of 1428-74i and 1514-7i  MeV, respectively.
{\color{blue}S}een only from the {\color{blue}perspective of} complex energy, pole $2$ might seem close to the $\bar{K}N$ threshold,
which makes counter-intuitive the fact that $2+2^\ast$ mainly contributes to the tail of the spectrum much above the $\bar{K}N$ threshold. 
Pole $2$, however, is not close to the $\bar{K}N$ threshold because it is positioned on Riemann sheet ($-\,-$), not ($-\,+$).
In Fig.\,\ref{pole_pos_z},  on the $z$-plane, one can immediately see that the physical domain closest to pole 2 is much above the $\bar{K}N$ threshold.
\par
In Tabs.\,\ref{tab:psm}-\ref{tab:scat}, the residues of the poles are presented, which contain the information of wave function and formation processes.
\par
\begin{table*}[htpb]
  \centering
  \resizebox*{\columnwidth}{!}{
  \begin{tabular}{cccc}
  \hline\hline
  &pole 1&pole 2&pole 3\\ \hline
  $z_n^{(3)}$       &0.5243+0.3159i$\pm$0.0062$\pm$0.0058i&1.6402-1.042i$\pm$0.0684$\pm$0.0904i&2.3227-0.0687i$\pm$0.0033$\pm$0.0031i\\ 
  $\sqrt{s}_n^{(3)}$&1.4203-0.0475i$\pm$0.0011$\pm$0.0015i&1.4283-0.074i$\pm$0.01$\pm$0.0037i&1.5138-0.0068i$\pm$0.0003$\pm$0.0003i\\ 
  \hline\hline
  \end{tabular}}
  \caption{Results for the pole positions by the uniformized Mittag-Leffler expansion with $m=3$. $z_n$ is the dimensionless pole position on the $z$-plane and $\sqrt{s}_n$ in units of GeV on the $\sqrt{s}$-plane.}
  \label{tab:pole}
\end{table*} 
\begin{table*}[htpb]
  \centering
  \resizebox*{\columnwidth}{!}{
  \begin{tabular}{cccc}
  \hline\hline
  $W$ (GeV)&pole 1&pole 2&pole 3\\ \hline
  1.95-2.05&-0.3486+0.3026i$\pm$0.0154$\pm$0.0149i&0.2487-0.122i$\pm$0.053$\pm$0.0342i&-0.0016-0.0029i$\pm$0.0013$\pm$0.0014i\\ 
  2.05-2.15&-0.3809+0.3245i$\pm$0.0156$\pm$0.0135i&0.1451-0.1877i$\pm$0.0442$\pm$0.0225i&-0.0175-0.0081i$\pm$0.0034$\pm$0.0023i\\ 
  2.15-2.25&-0.2662+0.1989i$\pm$0.0121$\pm$0.0096i&0.0294-0.0919i$\pm$0.028$\pm$0.0183i&-0.0108-0.0133i$\pm$0.0029$\pm$0.0021i\\ 
  2.25-2.35&-0.2539+0.208i$\pm$0.013$\pm$0.0106i  &0.0165-0.0339i$\pm$0.0318$\pm$0.0227i&0.0014-0.0122i$\pm$0.0023$\pm$0.0021i \\ 
  2.35-2.45&-0.2016+0.2142i$\pm$0.0131$\pm$0.0104i&0.0864-0.0442i$\pm$0.0306$\pm$0.0189i&-0.004-0.0105i$\pm$0.0021$\pm$0.0019i \\ 
  2.45-2.55&-0.1595+0.1369i$\pm$0.0097$\pm$0.008i &0.0423-0.0179i$\pm$0.0219$\pm$0.0151i&-0.0038-0.0091i$\pm$0.0018$\pm$0.0017i\\ 
  2.55-2.65&-0.1072+0.0925i$\pm$0.008$\pm$0.006i  &0.025-0.0066i$\pm$0.0169$\pm$0.0119i&-0.0043-0.0065i$\pm$0.0016$\pm$0.0014i\\ 
  2.65-2.75&-0.0891+0.057i$\pm$0.0065$\pm$0.0046i &0.0189+0.0133i$\pm$0.0139$\pm$0.01i&-0.0039-0.0062i$\pm$0.0014$\pm$0.0012i\\ 
  2.75-2.85&-0.0657+0.0466i$\pm$0.0056$\pm$0.0042i&0.0161-0.0066i$\pm$0.0115$\pm$0.008i&-0.0053-0.0051i$\pm$0.0013$\pm$0.0011i\\ 
  \hline\hline
  \end{tabular}}
  \caption{Results for the residues of the invariant-mass distributions of $\pi^+\Sigma^-$ in units of $\mu$b/GeV in nine bins of the center-of-mass energy, $W$, by the Uniformized Mittag-Leffler expansion with $m=3$.}
  \label{tab:psm}
\end{table*} 
\begin{table*}
  \centering
  \resizebox*{\columnwidth}{!}{
  \begin{tabular}{cccc}
      \hline\hline
      $W$ (GeV)&pole 1&pole 2&pole 3\\ \hline
      1.95-2.05&-0.2247+0.542i$\pm$0.0319$\pm$0.0262i &0.358-0.2978i$\pm$0.0864$\pm$0.0491i&-0.0013-0.0038i$\pm$0.0017$\pm$0.0017i\\ 
      2.05-2.15&-0.1119+0.7353i$\pm$0.035$\pm$0.0301i &0.0861-0.542i$\pm$0.0823$\pm$0.0456i&-0.0165-0.0155i$\pm$0.0035$\pm$0.0033i\\ 
      2.15-2.25&0.1962+0.4702i$\pm$0.02$\pm$0.0162i   &0.2154-0.1012i$\pm$0.0524$\pm$0.0325i&0.002-0.0171i$\pm$0.0027$\pm$0.0026i  \\ 
      2.25-2.35&0.0662+0.3112i$\pm$0.0144$\pm$0.0129i &0.1313-0.0568i$\pm$0.0374$\pm$0.0233i&0.0081+0.001i$\pm$0.0014$\pm$0.002i   \\ 
      2.35-2.45&-0.0017+0.3091i$\pm$0.0116$\pm$0.0116i&0.2839+0.0335i$\pm$0.0461$\pm$0.0327i&0.0028-0.0026i$\pm$0.0018$\pm$0.0016i \\ 
      2.45-2.55&-0.0119+0.2237i$\pm$0.009$\pm$0.0088i &0.2132+0.017i$\pm$0.0346$\pm$0.0236i&0.0004-0.006i$\pm$0.0014$\pm$0.0012i  \\ 
      2.55-2.65&-0.0189+0.1726i$\pm$0.0075$\pm$0.0073i&0.1377-0.0008i$\pm$0.0248$\pm$0.0162i&-0.0006-0.0038i$\pm$0.001$\pm$0.0011i \\ 
      2.65-2.75&-0.0123+0.1263i$\pm$0.0062$\pm$0.0055i&0.1136-0.0044i$\pm$0.02$\pm$0.0131i&-0.0029-0.0035i$\pm$0.001$\pm$0.0009i \\ 
      2.75-2.85&-0.0173+0.0932i$\pm$0.0055$\pm$0.005i &0.0859-0.0121i$\pm$0.016$\pm$0.0096i&-0.0021-0.0028i$\pm$0.0009$\pm$0.0007i\\ \hline\hline
  \end{tabular}}
  \caption{Results for the residues of the invariant-mass distributions of $\pi^-\Sigma^+$ in units of $\mu$b/GeV by the Uniformized Mittag-Leffler expansion with $m=3$.
}
  \label{tab:psp}
\end{table*} 
\begin{table*}
  \centering
  \resizebox*{\columnwidth}{!}{
  \begin{tabular}{cccc}
      \hline\hline
  $W$ (GeV)&pole 1&pole 2&pole 3\\ \hline
  1.95-2.05&-0.6515+0.3471i$\pm$0.2256$\pm$0.1211i&0.5316-1.2492i$\pm$0.7596$\pm$1.3581i &1.3537-0.6183i$\pm$2.7107$\pm$1.0427i \\ 
  2.05-2.15&-0.3179+0.5296i$\pm$0.0374$\pm$0.06i  &-0.3174-0.6043i$\pm$0.1764$\pm$0.1197i&-0.011+0.0019i$\pm$0.0104$\pm$0.0121i \\ 
  2.15-2.25&-0.1085+0.3535i$\pm$0.0209$\pm$0.0333i&-0.0763+0.0737i$\pm$0.1051$\pm$0.0997i&-0.0015-0.009i$\pm$0.0108$\pm$0.0099i \\ 
  2.25-2.35&-0.053+0.2798i$\pm$0.0154$\pm$0.0245i &0.0799+0.2387i$\pm$0.0854$\pm$0.0871i &0.0081-0.0087i$\pm$0.0086$\pm$0.0082i \\ 
  2.35-2.45&0.0027+0.2895i$\pm$0.0139$\pm$0.0227i &0.1853+0.2406i$\pm$0.0828$\pm$0.0885i &0.0052-0.001i$\pm$0.0079$\pm$0.0073i  \\ 
  2.45-2.55&0.0223+0.2323i$\pm$0.0097$\pm$0.0164i &0.1871+0.2054i$\pm$0.0618$\pm$0.0691i &-0.0038-0.0032i$\pm$0.0061$\pm$0.0063i\\ 
  2.55-2.65&0.0088+0.1641i$\pm$0.0084$\pm$0.0141i &0.1101+0.1044i$\pm$0.0479$\pm$0.0491i &-0.0051-0.0098i$\pm$0.0054$\pm$0.0042i\\ 
  2.65-2.75&-0.0018+0.1221i$\pm$0.0076$\pm$0.0126i&0.0883+0.1107i$\pm$0.0414$\pm$0.0428i &-0.0026-0.0058i$\pm$0.0047$\pm$0.0038i\\ 
  2.75-2.85&0.0089+0.094i$\pm$0.0058$\pm$0.009i   &0.0417+0.0439i$\pm$0.0317$\pm$0.0294i &0.0018-0.0052i$\pm$0.0025$\pm$0.0032i \\ \hline\hline
  \end{tabular}}
  \caption{Results for the residues of the invariant-mass distributions of $\pi^0\Sigma^0$ in units of $\mu$b/GeV  in nine bins of the center-of-mass energy, $W$, by the Uniformized Mittag-Leffler expansion with $m=3$.
}
  \label{tab:pso}
\end{table*}   
\begin{table*}
  \centering
  \resizebox*{\columnwidth}{!}{
  \begin{tabular}{cccc}
      \hline\hline
  &pole 1&pole 2&pole 3\\ \hline
       $K^-p\to K^-p$     &-5579+21810i$\pm$28869$\pm$8950i&6572-5272i$\pm$6311$\pm$4234i       &-99.33+32.89i$\pm$39.09$\pm$44.88i\\ 
  $K^-p\to\overline{K}^0n$&76090-9251i$\pm$25010$\pm$6306i &-1596+6936i$\pm$3447$\pm$3629i      &-188.2+64.44i$\pm$13.5$\pm$15.32i \\ 
  $K^-p\to\pi^+\Sigma^-$  &18960-96.75i$\pm$6251$\pm$1890i &-125.0+677.6i$\pm$1223.3$\pm$936.0i &-105.7+35.68i$\pm$7.2$\pm$8.39i   \\ 
  $K^-p\to\pi^-\Sigma^+$  &-4998+3449i$\pm$5546$\pm$1850i  &82.03+26.41i$\pm$1316.93$\pm$837.59i&-120.9+26.07i$\pm$8.7$\pm$9.86i   \\ \hline\hline
  \end{tabular}}
  \caption{Results for the residues of the cross sections, $K^-p\to K^-p$, $\bar{K}^0n$, $\pi^+\Sigma^-$, $\pi^-\Sigma^+$, in units $\mu$b/GeV$^2$ by the Uniformized Mittag-Leffler expansion with $m=3$.
}
  \label{tab:scat}
\end{table*} 
\clearpage
\subsection{Convergence of Uniformized Mittag-Leffler expansion from $m=1$ to $m=3$ }
\begin{figure}[!htb]
  \centering
  \includegraphics[width=\linewidth]{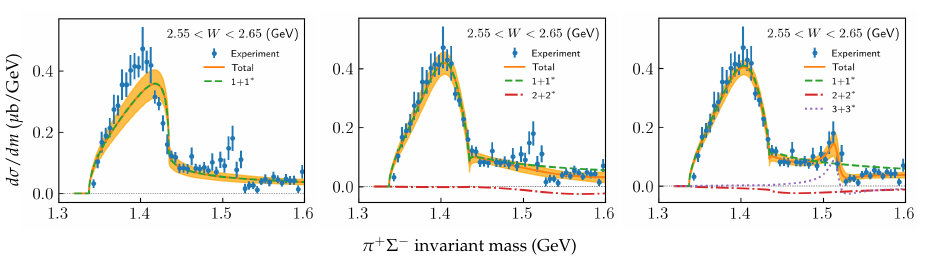}
  \caption{Results for the invariant-mass distribution of $\pi^+\Sigma^-$ in the reaction, $\gamma p \rightarrow K^+ \pi \Sigma$, by the Uniformized Mittag-Leffler expansion with $m=1$, $2$ and $3$ from left to right.
}
  \label{fig:m123_lineshape}
\end{figure}

Fig.\,\ref{fig:m123_lineshape} shows a typical invariant-mass distribution of $\pi^+\Sigma^-$ in the reaction, $\gamma p \rightarrow K^+ \pi \Sigma$, ($2.55 < W < 2.65$ (GeV)) from \cite{Moriya:2013eb}, fitted by the Uniformized Mittag-Leffler expansion in cases $m=1$, $2$ and $3$. 
As one can observe, the case $m=1$, fails to reproduce the broad peak structure below the $\bar{K}N$ threshold, whereas the case, $m=2$, reproduces most of the spectrum below and above the $\bar{K}N$ threshold. 
Comparing the cases, $m=1$ and $2$, it is clear that we need at least two resonant-pole pair contributions to successfully reproduce the broad peak structure below the $\bar{K}N$ threshold and the continuous spectrum above the $\bar{K}N$ threshold. 
By the addition of the third resonant-pole pair contribution, the narrow peak structure around 1520 MeV can also be taken into account, resulting in a satisfying approximation of the actual spectrum.
\par
\begin{figure}[!htb]
  \centering
  \includegraphics[width=0.7\linewidth]{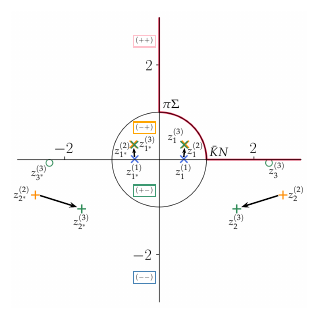}
  \caption{Results for the pole positions on the $z$-plane by the Uniformized Mittag-Leffler expansion with $m=1$, $2$ and $3$.
}
  \label{fig:m123_polz}
\end{figure}

Tab.\,\ref{tab:lam_pos_m123}, and Fig.\,\ref{fig:m123_polz} display the fitted pole positions for cases, $m=1$, $2$ and $3$.
The position of pole $1$ significantly shifts as we increase the number of terms from $m=1$ to $m=2$, whereas it hardly moves when increasing from $m=2$ to $m=3$.
This implies that the convergence of pole $1$ is almost realized for the case, $m=3$. 
The convergence of pole $2$ cannot be seen up to $m=3$
but pole $2$ and pole $3$ are positioned {\color{blue}further and further} away from pole $1$. 
\par
These behaviors imply that the expansion with $m=3$ is almost convergent in the vicinity of pole $1$.

\begin{table}[!htb]
  \centering 
  \begin{tabular}{cccc}
    \hline \hline
    &$m=1$&$m=2$&$m=3$\\ \hline
    $z_1^{(m)}$&0.52+0.012i$\pm$0.01$\pm$0.009i&0.551+0.323i$\pm$0.007$\pm$0.008i&0.524+0.316i$\pm$0.006$\pm$0.006i\\
    $\sqrt{s}_1^{(m)}$&1.478-0.003i$\pm$0.004$\pm$0.002i&1.420-0.042i$\pm$0.001$\pm$0.002i&1.420-0.048i$\pm$0.001$\pm$0.002i\\
    $z_2^{(m)}$&-&2.62-0.75i$\pm$0.09$\pm$0.06i&1.64-1.04i$\pm$0.07$\pm$0.09i\\
    $\sqrt{s}_2^{(m)}$&-&1.53-0.083i$\pm$0.01$\pm$0.004i&1.43-0.074i$\pm$0.01$\pm$0.004i\\ 
    $z_3^{(m)}$&-&-&2.323-0.069i$\pm$0.003$\pm$0.003i\\
    $\sqrt{s}_3^{(m)}$&-&-&1.5138-0.0068i$\pm$0.0003$\pm$0.0003i\\ \hline \hline
  \end{tabular}
  \caption{Results for the pole positions by the Uniformized Mittag-Leffler expansion with $m=1$, $2$ and $3$.  $z_n$ $(n=1,2,3)$ is the dimensionless pole position on the $z$-plane and $\sqrt{s}_n$ $(n=1,2,3)$ is the pole position on the $\sqrt{s}$-plane with units of GeV.}
  \label{tab:lam_pos_m123}
\end{table}
\par
\subsection{Discussion}
As stated above, we found only {\color{blue} a single} pair of poles, $1 + 1^*$, on the ($-\,+$) sheet of the complex $\sqrt{s}$ plane, which {\color{blue}sufficiently} explains the broad peak structure between the $\pi\Sigma$ and $\bar K N$ thresholds.
Its contribution to the Unformized Mittag-Leffler exapansion converges up to $m=3$.
This leads us to identify Pole $1$ as $\Lambda(1405)$.
Also, it is natural to identify Pole $3$ as  $\Lambda(1520)$ due to its small width,
even though the convergence of $3 + 3^*$ has not been confirmed up to $m=3$.
The interpretation of Pole $2$ is less intuitive, which cannot be identified {\color{blue} with any} physical resonance.
The contribution of $2 + 2^*$ gives {\color{blue}the} continuous spectrum above the $\bar K N$ threshold together with the tail of the contribution of  $1 + 1^*$.
It should be also noted that the contribution of $2 + 2^*$ is mostly negative.
\par
Usually, the observed spectrum is naively interpreted as the sum of physical resonances and background contributions.
However, there is no well-defined criterion when a pole should be identified as a physical resonance or not.
In the Uniformized Mittag-Leffler expansion, the observed spectrum is represented as a sum of pole contributions, which is well defined.
There is no need to identify a pole as a physical resonance or not.
Obviously, all the pole contributions in the Uniformized Mittag-Leffler expansion cannot be interpreted as resonance contributions in the usual sense.
\par
The results obtained in a model-independent manner by the use of the Uniformized Mittag-Leffler expansion support a single-pole picture of $\Lambda(1405)$.
In order to {\color{blue}solidify this claim,} it may be useful to take into account more than three resonant-pole pair terms in the Uniformized Mittag-Leffler expansion.
\section{Summary and Conclusion}
In this paper we applied the Uniformized Mittag-Leffler expansion, proposed in our previous paper, to {\color{blue}the} $\Lambda(1405)$ {\color{blue} resonance}.
We expanded the observable as a sum of resonant-pole pairs {\color{blue}with} a variable which expresses the observable {\color{blue}to be} single-valued, and fitted it to experimental data {\color{blue}of} the invariant-mass distribution of $\pi^+\Sigma^-$, $\pi^-\Sigma^+$, $\pi^0\Sigma^0$ final states in the reaction, $\gamma p \rightarrow K^+ \pi \Sigma$, and the elastic and inelastic cross sections, $K^-p\to K^-p$, $\bar{K}^0n$, $\pi^+\Sigma^-$, $\pi^-\Sigma^+$.
Thus, we determined the resonant energy, width and residues in a model-independent manner.
\par
We started from one pair and {\color{blue}gradually} increased the number of pairs up to three.
We observed that the first pair converges while the second and third pairs emerge {\color{blue}further and further} away from the first pair,
which implies that the Uniformized Mittag-Leffler expansion with three pairs is almost convergent in the vicinity of $\Lambda(1405)$.
The reduced chi square values are 5.74, 2.65 and 1.18 for the number of pairs, one, two and three, respectively,
and the Uniformized Mittag-Leffler expansion with three pairs satisfactorily fits experimental data.
The broad peak structure between the $\pi\Sigma$ and $\bar K N$ thresholds regarded to be $\Lambda(1405)$ is explained by the first pair,
while the continuous spectrum above the $\bar{K}N$ threshold is given by the first and second pairs except for the narrow structure around 1520 MeV, which is explained by the third pair.
The results are consistent with the single-pole picture of  $\Lambda(1405)$
with a resonant energy of 1420 $\pm$ 1 MeV, and {\color{red} a half} width of 48 $\pm$ 2 MeV.
\par
In conclusion, the Uniformized Mittag-Leffler Expansion approach turns out to be very powerful.
If experimental data have enough statistics, one can determine the information of near-threshold resonances in a model-independent way.
This is extremely important in order to achieve unbiased understanding of near-threshold resonances.
\par
As an extension of the present work, we can {\color{blue}procede in} two directions.
One is the application of the present Uniformized Mittag-Leffler expansion to other hadron resonances, which are {\color{blue}positioned} near two two-body thresholds.
The other is the extension of the present Uniformized Mittag-Leffler expansion {\color{blue}to} the case {\color{blue}with three or} more two-body thresholds {\color{blue}or cases with} three-body thresholds.
Both {\color{blue}possibilities} are {\color{blue}presently} under our consideration.
\par
\clearpage
\bibliography{paper_lambda1405}
\end{document}